\documentstyle[12pt]{article}

\newcommand{\reseteqnum}{\setcounter{equation}{0}}
\newcommand{\Half}{\frac{1}{2}}

\newcommand{\vev}[1]{\left\langle #1\right\rangle}
\newcommand{\e}{\epsilon}

\newcommand{\wt}[1]{\widetilde{#1}}

\newcommand{\rd}[1]{\overleftarrow{\partial}_{\!\! #1}}
\newcommand{\ld}[1]{\overrightarrow{\partial}_{\!\! #1}}
\newcommand{\p}{\partial}
\newcommand{\ed}{{\rm d}}
\newcommand{\Diff}[2]{\frac{\partial #1}{\partial #2}}
\newcommand{\diff}[2]{\partial #1/\partial #2}
\newcommand{\DDiff}[3]{\frac{\partial^2 #1}{\partial #2\partial #3}}

\newcommand{\rFDiff}[1]{\frac{\overleftarrow{\delta}}{\delta #1}}
\newcommand{\lFDiff}[1]{\frac{\overrightarrow{\delta}}{\delta #1}}
\newcommand{\FDiff}[2]{\frac{\delta #1}{\delta #2}}

\newcommand{\calM}{{\cal M}}
\newcommand{\calO}{{\cal O}}
\newcommand{\calA}{{\cal A}}
\newcommand{\calB}{{\cal B}}

\newcommand{\calL}{{\cal L}}
\newcommand{\calD}{{\cal D}}
\newcommand{\calZ}{{\cal Z}}
\newcommand{\B}[1]{\overline{#1}}
\newcommand{\AB}[2]{\left(#1,#2\right)}
\newcommand{\DELTA}{\kern1pt\vbox{\hrule height .6pt
            \hbox{\vrule width .6pt\hskip 3pt
            \vbox{\vskip 8pt}\hskip 5pt\vrule width .6pt}
            \hrule height .6pt}\kern1pt}
\newcommand{\Ngh}{N_{gh}}
\newcommand{\VEV}[1]{\langle\!\langle #1 \rangle\!\rangle}
\newcommand{\invL}{\frac{1}{\lambda}}
\newcommand{\dB}{\delta_{\rm B}}
\newcommand{\hhdB}{\kern1.5pt\widehat{\kern-1.5pt\widehat \delta}_{\rm B}}
\newcommand{\hhI}{\kern3pt\widehat{\kern-3pt\widehat I}}
\newcommand{\hhV}{\widehat{\widehat V}}

\begin{document}

\renewcommand{\thepage}{ }
\begin{titlepage}
\title{
\hfill
\parbox{4cm}{\normalsize KUNS-1212\\HE(TH)~93/08\\hep-th/9308001}\\
\vspace{1cm}
``Theory of Theories'' Approach\\ to String Theory}
\author{Hiroyuki Hata\thanks{e-mail address:
hata@gauge.scphys.kyoto-u.ac.jp, hata@jpnyitp.bitnet}
{\,}\thanks{Supported in part by Grant-in-Aid for Scientific Research
from Ministry of Education, Science and Culture
(\# 05230037).}\\
{\normalsize\em Department of Physics, Kyoto University}\\
{\normalsize\em Kyoto 606-01, Japan}}
\date{\normalsize August, 1993}

\maketitle

\begin{abstract}
\normalsize
We propose a new formulation of gauge theories as a quantum theory
which has the gauge theory action $S$ as its dynamical variable.
This system is described by a simple actional $I(S)$ (that is, an
action for the action $S$) whose equation of motion gives the
Batalin-Vilkovisky (BV) master equation for $S$.
Upon quantization we find that our new formulation is reduced to
something like a topological field theory having a BRST exact
gauge-fixed actional. Therefore the present formulation can reproduce
ordinary gauge theories since the path-integral over $S$ is dominated
by the classical configuration which satisfies the BV master equation.
This ``theory of theories'' formulation is intended to be applied to
closed string field theory.
\end{abstract}
\end{titlepage}

\newpage
\renewcommand{\thepage}{\arabic{page}}
\setcounter{page}{1}
\baselineskip=17pt plus 0.2pt minus 0.1pt
\parskip=7pt

\section{Introduction}
\reseteqnum

In this paper we propose a new formulation of gauge field theory.
It applies in principle to any kind of gauge theory, but our
interest is mainly in the reformulation of string field theory.

A field theory having a dynamical variable $\varphi$ is described by
an action $S(\varphi)$ as
\begin{equation}
\vev{A}=\int\!\calD\varphi\, A(\varphi)\exp S(\varphi) .
\label{eq:theory}
\end{equation}
In gauge theories $S$ must be modified so that the local gauge
invariance is fixed and correspondingly the Faddeev-Popov (FP) ghosts
are introduced. This procedure for quantizing gauge theories is most
efficiently carried out using the BRST or the Batalin-Vilkovisky (BV)
\cite{BV} formalism. According to the BV formalism the quantum action
$S$ has to satisfy the master equation
\begin{equation}
\hbar\Delta S + \Half\{S,S\}= 0 ,\label{eq:mastereq}
\end{equation}
where $\{*,*\}$ and $\Delta$ are operators whose precise definitions
will be given in section 2. The master equation expresses the quantum
BRST invariance of the system: the term proportional to $\hbar$
takes into account the variation of the path-integral measure under
the BRST transformation. (Precisely speaking, the action $S$ in
eq.~(\ref{eq:theory}) for a gauge theory is obtained from $S$
satisfying the master equation (\ref{eq:mastereq}) by the
restriction to a Lagrangian submanifold. In this section we do not
distinguish these two $S$'s to avoid unnecessary complication.)

In ``simple'' systems such as the Yang-Mills theory, the measure
term $\hbar\Delta S$ can be consistently neglected by using dimensional
regularization. However, for closed string field theory, which is
recognizable as a gauge theory having an infinite number of gauge
symmetries, the measure term is essential in obtaining a consistent
theory. For such a system the quantum action $S$ satisfying
eq.~(\ref{eq:mastereq}) is given as an infinite power series in
$\hbar$:
\begin{equation}
S=\sum_{n=0}^\infty \hbar^n S^{(n)} .\label{eq:sumS}
\end{equation}
Construction of the quantum action (\ref{eq:sumS}) for closed
string field theory has been carried out in
refs.~\cite{Hata}, \cite{Zwiebach} and \cite{Zwiebach-long}.
Unfortunately, the resulting $S$ looks too complicated to be used in
the investigation of (possible) non-perturbative aspects of string
theory. Invention of another, much simpler reformulation of string
(field) theory is greatly desired.

Our attempt in this paper is to present such a reformulation of gauge
theories, and in particular of closed string field theory, without
referring to the explicit form of the action $S(\varphi)$.
Instead we promote the action $S(\varphi)$ from a fixed functional of
$\varphi$ to a {\em dynamical variable} which should be
path-integrated out.
Since the kind of field $\varphi$ which may be used as the argument of
$S$ is fixed, the coupling constants in $S(\varphi)$ may be regarded
as the dynamical variables.
Roughly speaking, we consider a theory described by the path-integral
\begin{equation}
\VEV{\calO}=\int\!\calD S\,\calO(S)
\exp\left(\frac{1}{\lambda}I(S)\right) , \label{eq:roughly}
\end{equation}
where $I(S)$ is the action for the action (hereafter called the
{\em actional}). Since $S(\varphi)$ specifies a theory, the present
formulation may be called a ``Theory of theories'' (TT).

The principles we use in constructing the actional $I(S)$ for TT are
as follows. First, we require that the equation of motion of TT,
$\delta I(S)/\delta S=0$, gives the master equation
(\ref{eq:mastereq}). Second, $I(S)$ should be invariant under
the ``local'' gauge transformation
\begin{equation}
\delta_\e S = \Delta \e + \{S,\e\} , \label{eq:gaugetransf}
\end{equation}
where $\e$ is an arbitrary functional of $\varphi$. The transformation
(\ref{eq:gaugetransf}) is known to be a symmetry of the master
equation (\ref{eq:mastereq}) \cite{Witten-ab,HT}: if $S$ is a solution
to the BV equation, so is $S+\delta\e S$ (this can be naively
understood if the LHS of the BV equation (\ref{eq:mastereq}) is
regarded as an analogue of the usual field strength $F=\ed A+A^2$).

An actional $I(S)$ satisfying the above two requirements is easily
found (and has already been proposed in ref.~\cite{HataZwiebach}).
We want TT to reproduce the original gauge theory (\ref{eq:theory}),
since our aim is to present a reformulation of string field theory.
This implies that our TT should be a kind of ``topological'' theory
\cite{Witten-TFT,BBRT}
which has almost no physical degrees of freedom as a system of the
dynamical variable $S(\varphi)$ (recall that an ordinary gauge theory
(\ref{eq:theory}) is described by a {\em fixed} action $S(\varphi)$).
Remarkably we find that this expectation is in fact true.
Since TT is also a gauge theory, we quantize it by again employing the
BV formalism. After introducing an auxiliary field, the resulting
quantized TT turns out to be a topological theory described by a BRST
exact actional. The connection between TT and the conventional
formulation of gauge theories is made through the partition
function: in TT the partition function operator $V_L(S)$ is an
observable, whose expectation value is shown to be equal to the
partition function of a gauge theory in the conventional formulation.

The organization of the rest of this paper is as follows. In section 2
we give a brief summary of the BV formalism necessary in the
construction of TT. In section 3, which is the main part of this paper,
we first introduce the actional (sect.\ 3.1), carry out the BV
quantization of TT (sect.\ 3.2, 3.3 and 3.4), and discuss the
relationship to the ordinary  formulation of gauge theories
(sect.\ 3.5). The final section (sect.\ 4) is devoted to a summary and
discussion.

\section{BV formalism}
\reseteqnum

In this section we shall recapitulate the elements of the BV formalism
used in this paper. We follow our previous convention
\cite{HataZwiebach}. A more detailed explanation of the BV formalism
may be found in ref.~\cite{HT}.

We consider an $(n,n)$-dimensional supermanifold $\calM$.
The coordinates of $\calM$ are the field variables.
In real gauge theories, a field $\varphi(x)$ has continuous space-time
parameter $x$. Here the index $I\ (=1,\cdots,2n)$ specifying the
coordinates of $\calM$ should be understood to represent all the
(continuous as well as discrete) parameters characterizing the fields.

The supermanifold $\calM$ is endowed with an odd symplectic structure
defined by a fermionic two-form $\omega$ which is non-degenerate and
closed, $\ed\omega=0$, and carries the ghost number $\Ngh[\omega]=+1$.
In a local coordinate system $(z^I)=(z^1,z^2,\ldots,z^{2n})$ of
$\calM$, $\omega$ is expressed as
\begin{equation}
\omega= -\ed z^I\omega_{IJ}(z)\,\ed z^J
= \omega_{JI}(z)\,\ed z^I \wedge\ed z^J .
\label{eq:omega}
\end{equation}
On $\calM$ we also introduce the volume element
\begin{equation}
d\mu(z)=\rho(z)\prod_{I=1}^{2n}\ed z^I , \label{eq:dmu}
\end{equation}
where $\rho(z)$ is the density.
Then we can define two basic operators, the antibracket $\{*,*\}$ and
the delta-operator $\Delta_\rho$, by
\begin{eqnarray}
&&\{A,B\}=A\rd{I}\,\omega^{IJ}(z) \ld{J}B ,\label{eq:antibracket}\\
\noalign{\vskip.2cm}
&&\Delta_\rho A ={\displaystyle \frac{1}{2\rho}}(-)^{\displaystyle z^I}
\p_I\!\left(\rho\,\omega^{IJ}\p_J A\right) ,\label{eq:Delta}
\end{eqnarray}
where $\omega^{IJ}(z)$ is the inverse matrix to $\omega_{IJ}(z)$,
and $\ld{I}=\p_I=\p_l/\p z^I$ and $\rd{I}=\p_r/\p z^I$ denote the
left- and right-derivatives respectively.\footnote{
$(-)^A=+1$ ($-1$) if $A$ is Grassmann even (odd).}
Note that both the antibracket and $\Delta$ raise the ghost number
$\Ngh$ by one.

The antibracket and the delta-operator satisfy the three basic
properties:
\begin{eqnarray}
&\left(\Delta\right)^2 = 0 ,
&\mbox{(nilpotency)}\label{eq:nilpotency}\\
&\Delta\{A,B\}=\{\Delta A,B\}+(-)^{A+1}\{A,\Delta B\} ,
&\mbox{(Leibniz rule)} \label{eq:Leibniz}\\
&(-)^{(A+1)(C+1)}\left\{\{A,B\},C\right\} + \mbox{cyclic}(A,B,C)=0 ,
&\mbox{(Jacobi identity)} .\label{eq:Jacobi}
\end{eqnarray}
Eqs.~(\ref{eq:Leibniz}) and (\ref{eq:Jacobi}) are consequences
of $\ed\omega=0$, while eq.~(\ref{eq:nilpotency}) is a requirement on
the density $\rho(z)$\footnote{
As a matter of fact, $\ed\omega=0$ follows from $\Delta^2=0$.}.
Other useful formulas concerning the antibracket and the
delta-operator are \cite{Witten-ab,HT}
\begin{equation}
(-)^A\{A,B\}=\Delta\!\left(A B\right)-
\Delta A\cdot B - (-)^A A\Delta B , \label{eq:Witten-formula}
\end{equation}
and
\begin{eqnarray}
&&\{A,B\}=-(-)^{(A+1)(B+1)}\{B,A\} , \label{eq:AB1}\\
&&\{A,BC\}=\{A,B\}C + (-)^{(A+1)B}B\{A,C\} ,\label{eq:AB2}\\
&&\{AB,C\}=A\{B,C\} + (-)^{B(C+1)}\{A,C\}B \label{eq:AB3} .
\end{eqnarray}

The master equation for $S(z)$ reads
\begin{equation}
M(S)\equiv\Delta S + \Half\{S,S\}=0 ,\label{eq:BV}
\end{equation}
or equivalently
\begin{equation}
\Delta e^S = 0 .\label{eq:BVe}
\end{equation}
Given a $S(z)$ satisfying the master equation (\ref{eq:BV}) the
gauge-fixed quantum theory is defined by the path-integral
\begin{equation}
\vev{A}=\int_L\! d\lambda\,A(z)\exp S(z) ,\label{eq:vevA}
\end{equation}
where the integration is over the Lagrangian submanifold $L$ and
$d\lambda$ is the associated integration measure.
The Lagrangian submanifold $L$ is a $(k,n-k)$-dimensional submanifold
of $\calM$, such that $\omega(v,\tilde v)=0$ for any pair, $v$ and
$\tilde v$, of tangent vectors to $L$ at $z\in L$
($v,\tilde v\in T_zL$).
The corresponding volume element $d\lambda$ is defined by
\begin{equation}
d\lambda(e_1,\cdots,e_n)=d\mu(e_1,\cdots,e_n,f^1,\ldots,f^n)^{1/2} ,
\label{eq:dlambda}
\end{equation}
where $d\mu$ is the volume element in $\calM$, eq.~(\ref{eq:dmu}), and
$(e_1,\cdots,e_n,f^1,\ldots,f^n)$ is a basis of the tangent space
$T_z\calM$ such that $(e_1,\cdots,e_n)$ is a basis of $T_zL$ and the
condition $\omega(e_i,f^j)=\delta_i^j$ is satisfied.
The choice of $L$ corresponds to the choice of gauge fixing.
In order for the expectation value $\vev{A}$ (\ref{eq:vevA}) to be
independent of the choice of the Lagrangian submanifold $L$, the
operator $A(z)$ has to satisfy the following condition (see below):
\begin{equation}
\Delta A + \{S,A\}=0 .\label{eq:observable}
\end{equation}

The solution $S(z)$ of the master equation (\ref{eq:BV}) is not
unique. Given a solution $S(z)$, we have a continuous family of
solutions obtained by the infinitesimal ``gauge transformation''
$\delta_\e$ \cite{Witten-ab,HT}:
\begin{equation}
\delta_\e S = \Delta \e + \{S,\e\} ,\label{eq:de}
\end{equation}
where the transformation parameter $\e(z)$ carries $\Ngh[\e]=-1$.
In fact, using eqs.~(\ref{eq:nilpotency})--(\ref{eq:Jacobi}), the
master equation $M(S)$ of eq.~(\ref{eq:BV}) is shown to transform
homogeneously under $\delta_\e$: $\delta_\e M(S)=\{M(S),\e\}$.
The same formulas tell that the transformation $\delta_\e$ forms a
closed algebra:
\begin{equation}
\left[\delta_{\e_1},\delta_{\e_2}\right]=\delta_{\{\e_1,\e_2\}} .
\label{eq:algebra}
\end{equation}
The finite transformation which has $\delta_\e$ of eq.~(\ref{eq:de})
as its infinitesimal expression is given by considering a general
canonical transformation $g: \calM \rightarrow \calM$ satisfying
$g^*\omega=\omega$ (see ref.~\cite{HataZwiebach}).

The relationship between the master equation (plus the requirement
(\ref{eq:observable}) on $A$) and the independence of
eq.~(\ref{eq:vevA}) on the choice of $L$ is understood as follows.
A general infinitesimal deformation of $L$ may be realized by a canonical
transformation: $z^I \rightarrow z^I + \{z^I,\e\}$ for some $\e(z)$.
Therefore, we have
\begin{eqnarray}
&\vev{A}_{L+\delta L} - \vev{A}_L
&= \int_L d\lambda\left(\Delta\e\cdot A e^S + \{A e^S,\e\}\right)
\nonumber \\
&&= \int_Ld\lambda\left(\Delta\!\left(\e\, A e^S\right)
    + \e\,\Delta\!\left(A e^S\right)\right) ,\label{eq:deltaL}
\end{eqnarray}
where the $\Delta\e$ term in the first expression originates from the
change of the measure $d\lambda$, while the $\{A e^S,\e\}$ term
expresses the coordinate transformation on the integrand
$A(z)e^{S(z)}$ (see section 3.3 of ref.~\cite{HataZwiebach}).
The last expression of eq.~(\ref{eq:deltaL}) is obtained by using the
formula (\ref{eq:Witten-formula}).
The $\Delta\!\left(\e Ae^S\right)$ term in the last expression of
eq.~(\ref{eq:deltaL}) vanishes due to the general formula \cite{Schwarz}
\begin{equation}
\int_L d\lambda\,\Delta A = 0 ,\label{eq:Schwarz-th}
\end{equation}
which holds for arbitrary $A(z)$. The master equation and the
condition (\ref{eq:observable}) ensure the vanishing of
$\Delta\!\left(A e^S\right)=\left(\Delta A +\{S,A\}\right)e^S +
A\Delta e^S$.

The simplest coordinate system for $\calM$ is the Darboux frame
$(\phi^i,\phi^*_i)_{i=1,\cdots,n}$ with
$\omega=-2\sum_i\ed\phi^i\wedge\ed\phi^*_i$.
The coordinates $\phi^i$ and $\phi^*_i$ are called fields and
antifields, respectively, and they satisfy
$\Ngh[\phi^i]+\Ngh[\phi^*_i]=-1$.
The antibracket and the delta-operator in the Darboux frame (with
$\rho(z)=1$) are given by
\begin{eqnarray}
&&\{A,B\}=\Diff{_r A}{\phi^i}\Diff{_l B}{\phi^*_i} -
\Diff{_r A}{\phi^*_i}\Diff{_l B}{\phi^i} ,\label{eq:antibracketinDB}\\
&&\Delta = (-)^{\phi^i}\DDiff{_l}{\phi^i}{\phi^*_i} .\label{eq:DeltainDB}
\end{eqnarray}
In the Darboux frame the Lagrangian submanifold $L$ is specified by
\begin{equation}
L: \phi^*_i=\Diff{\Upsilon(\phi)}{\phi^i} , \label{eq:LinDB}
\end{equation}
where $\Upsilon(\phi)$ is called gauge fermion
($\Ngh[\Upsilon]=-1$).
The expectation value of an observable $A$, eq.~(\ref{eq:vevA}),
now reads
\begin{equation}
\vev{A}=\int\! d\phi\, A(\phi,\phi^*) \exp S(\phi,\phi^*)
\Big\vert_{\,\phi^*_i=\diff{\Upsilon(\phi)}{\phi^i}} .
\label{eq:vevAinDB}
\end{equation}
Finally we shall explain the BRST transformation in the BV quantized
theory (\ref{eq:vevAinDB}) in the Darboux frame.
First we define the pre-BRST transformation $\dB$ on a general $A(z)$
by
\begin{equation}
\dB A = \{A,S\} . \label{eq:preBRSTforT}
\end{equation}
Then the BRST transformation in the gauge-fixed theory $\widehat\dB$
is defined by restricting $\dB$ to the Lagrangian submanifold:
\begin{equation}
\widehat\dB\phi^i = \dB\phi^i\vert_L = \Diff{S(\phi,\phi^*)}{\phi^*_i}
\bigg\vert_{\,\phi^*_i=\diff{\Upsilon(\phi)}{\phi^i}} . \label{eq:hatdBphi}
\end{equation}
The master equation ensures the quantum BRST invariance and the
on-shell nilpotency of $\widehat\dB$:
\begin{eqnarray}
&&\widehat\dB\left( \widehat S(\phi) + \ln\prod_i d\phi^i\right)=0 ,
\label{eq:quntumBRSTinvofT}\\
&&\left(\widehat\dB\right)^2\phi \propto
\FDiff{\widehat S(\phi)}{\phi} ,\label{eq:onshellNPofT}
\end{eqnarray}
where $\widehat S(\phi)\equiv S(\phi,\phi^*=\diff{\Upsilon}{\phi})$.

\section{Theory of theories}
\reseteqnum

As seen above, the master equation has a local gauge symmetry
given by $\delta_\e$ of eq.~(\ref{eq:de}).
In this section, we shall construct TT, namely, a gauge theory
which has $S(z)$ as its dynamical variable and and has an invariance
under $\delta_\e$. We shall then study the quantization of TT based
on the BV formalism and the relationship to the ordinary formulation
on gauge theories.

\subsection{Actional}

First we need the actional $I(S)$ for TT. As stated in section 1, we
demand that $I(S)$ has an invariance under $\delta_\e$ and that the
equation of motion, $\delta I(S)/\delta S=0$, gives the master equation
(\ref{eq:BV}).
Such an actional has been proposed in ref.~\cite{HataZwiebach}.
It takes a fairly simple form:
\begin{equation}
I_{hz}(S)=\int_\calM d\mu H(z)\Delta H(z) ,\label{eq:calA}
\end{equation}
where $H(z)\equiv\exp S(z)$ (see section 3.2 of
ref.~\cite{HataZwiebach}, where we denoted $I_{hz}(S)$ by $\calA(S)$).
In this paper we adopt, instead of (\ref{eq:calA}), the slightly
modified actional $I(S)$:
\begin{equation}
I(S)=\Half\int_\calM d\mu(z) H(\B{z})\Delta H(z)
=\Half\int_\calM d\mu(z) H(z)\Delta H(\B{z}) ,\label{eq:I}
\end{equation}
where the coordinate $\B{z}^I$ is related to the original one $z^I$ by the
``inversion'' of the Grassmann odd components:
\begin{equation}
\B{z}^I \equiv (-)^{\displaystyle z^I}z^I . \label{eq:Bz}
\end{equation}
The two expressions of eq.~(\ref{eq:I}) are equivalent on using the
partial integration formula
\begin{equation}
\int_\calM d\mu A\Delta B=-\Half\int_\calM d\mu\{A,B\}
=(-)^A\int_\calM d\mu\left(\Delta A\right)B .
\label{eq:DeltaPI}
\end{equation}

Since we have $\Ngh[\Delta]=+1$ and $\Ngh[S(z)]=0$, the requirement that
the actional $I(S)$ carries no ghost number $\Ngh$ leads to the
following restriction on the ghost numbers of $z^I$:
\begin{equation}
\Ngh[d\mu(z)]\equiv
\sum_{I=1}^{2n}(-)^{\displaystyle z^I}\!\Ngh[z^I]=-1 .
\label{eq:Nghdmu}
\end{equation}
Since $\Ngh[z^I]=\mbox{even}$ (odd) if $z^I$ is Grassmann even (odd),
the requirement (\ref{eq:Nghdmu}) tells that the dimension $n$ of our
$(n,n)$ supermanifold $\calM$ must be an odd integer (this fact
follows immediately from the requirement that $I(S)$ be a bosonic
quantity since $\Delta$ is fermionic).
It is a delicate matter whether condition (\ref{eq:Nghdmu}) is
satisfied for a concrete system, such as string field theory,
since the index $I$ is in fact a continuous parameter.
Here we simply assume that the condition (\ref{eq:Nghdmu}) is
satisfied.

It is obvious that the equation of motion, $\delta I(S)/\delta S(z)=0$,
gives the master equation (\ref{eq:BVe}).
The invariance of $I(S)$ (\ref{eq:I}) under the gauge transformation
$\delta_\e$ of eq.~(\ref{eq:de}), which is expressed on $H(z)$ as
\begin{equation}
\delta_\e H = \Delta\!\left(H\e\right)-\left(\Delta H\right)\e ,
\label{eq:deH}
\end{equation}
is shown as follows:
\begin{eqnarray}
&\delta_\e I(S)&=\int\! d\mu \Bigl(\Delta\!\left(H \e\right)
- \Delta H\cdot \e\Bigr)\Delta\B{H} \nonumber\\
&&=\int\! d\mu\left( -H\Delta^2\B{H} + \Delta H\cdot\Delta\B{H}\right)\e
= 0 ,\label{eq:deI}
\end{eqnarray}
where $\B{H}$ is short for $H(\B{z})$, and we have used
eq.~(\ref{eq:DeltaPI}) and the nilpotency of $\Delta$.
The vanishing of the $\Delta H\Delta\B{H}\e$ term is understood by
making the change of variables from $z$ to $\B{z}$:
\begin{equation}
\int\! d\mu \Delta H\cdot\Delta\B{H}\cdot\e
=\int\!\left(-d\mu\right)\Delta\B{H}\cdot\Delta H\cdot(-\e) = 0 ,
\label{eq:DHDH}
\end{equation}
where we have used the properties
\begin{eqnarray}
&&d\mu(\B{z})= - d\mu(z) ,\label{eq:Bdmu}\\
&&\Delta(\B{z})= -\Delta(z) ,\label{eq:BDelta}\\
&&\e(\B{z})=-\e(z) . \label{eq:Be}
\end{eqnarray}
Eq.~(\ref{eq:Be}) is the restriction on $\e(z)$ that it should not
contain Grassmann odd ``constants''.
Note that $f(\B{z})=\pm f(z)$ depending on whether the function
$f(z)=\omega^{IJ}(z)$, $\rho(z)$ etc.\ is Grassmann even (upper sign) or
odd (lower sign). For example, we have $\rho(\B{z})=\rho(z)$.

\subsection{Master equation for TT}

Having presented the gauge invariant actional of TT, our next task is
to quantize it. We shall carry out this quantization using the BV
formalism.
For this purpose we have to first define antibracket and
delta-operator for TT, which we denote by $(*,*)$ and $\DELTA$,
respectively. Here we shall adopt the following ones:
\begin{eqnarray}
&&\AB{\calA}{\calB}
=\calA(S)\rFDiff{H(z)}\int\frac{d\mu(z)}{\left(\rho(z)\right)^2}
\cdot\lFDiff{H(\B{z})}\calB(S) \nonumber\\
&&\qquad\quad=\int\frac{d\mu(z)}{\left(\rho(z)\right)^2}
\cdot\lFDiff{H(z)}\calA(S)\cdot\lFDiff{H(\B{z})}\calB(S) ,\label{eq:AB}\\
\noalign{\vskip.2cm}
&&\DELTA\calA=\Half\int\frac{d\mu(z)}{\left(\rho(z)\right)^2}
\cdot\lFDiff{H(z)}\cdot\lFDiff{H(\B{z})}\calA(S) ,\label{eq:BOX}
\end{eqnarray}
where $\calA(S)$ and $\calB(S)$ are arbitrary functionals of $S(z)$.
Note that the differential operator $\delta/\delta H(z)$ is
Grassmann odd since we have
\begin{equation}
\lFDiff{H(z)}H(z')=-H(z')\rFDiff{H(z)}=\delta(z-z')=-\delta(z'-z) ,
\label{eq:FDiff}
\end{equation}
and $\delta(z-z')$ is Grassmann odd (recall that $n$ is odd).
In obtaining the second expression of eq.~(\ref{eq:AB}) we have used
the formula
\begin{equation}
\calA(S)\rFDiff{H(z)}=(-)^{\calA+1}\lFDiff{H(z)}\calA(S) .\label{eq:rlFDiff}
\end{equation}
It is easily seen that the antibracket $\AB{*}{*}$ and the ``delta''
operator $\DELTA$ satisfy the basic properties of
eqs.~(\ref{eq:nilpotency}) -- (\ref{eq:AB3})
with $\{*,*\}$ and $\Delta$ replaced with $\AB{*}{*}$ and $\DELTA$,
respectively.
Since $\Ngh[\delta/\delta H(z)]=\Ngh[\delta(z)]=-\Ngh[d\mu(z)]$, the
condition (\ref{eq:Nghdmu}) ensures that both $(*,*)$ and $\DELTA$
raise $\Ngh$ by one.

Here we should add a comment on the ghost number restriction on $S(z)$
as the argument of $I(S)$. Let us suppose a (formal) expansion of
$S(z)$ in terms of a ``complete set of interactions'' $\{f_n(z)\}$:
\begin{equation}
S(z)=\sum_{n}f_n(z)s_n ,\label{eq:expandS}
\end{equation}
where the coupling constants $s_n$ are now the dynamical variables.
When we consider the gauge invariant actional $I(S)$ (\ref{eq:I}) and
the gauge transformation (\ref{eq:de}), we can consistently restrict
the summation (\ref{eq:expandS}) to those $n$ with $\Ngh[s_n]=0$ by
restricting also the expansion of the transformation parameter
$\e(z)=\sum_n f_n(z)\e_n$ to $\Ngh[\e_n]=0$ (and therefore
$\Ngh[f_n]=-1$).
In other words, we can impose the condition
\begin{equation}
S(z_\theta)=S(z),\quad \e(z_\theta)=e^{-\theta}\e(z),
\quad (z_\theta^I\equiv e^{\theta\Ngh[z^I]}z^I)
\label{eq:restrictS}
\end{equation}
for an arbitrary $\theta$.
This restriction looks natural if we regard $I(S)$ as a ``classical''
actional before introducing the FP ghosts for quantization.
However, when we discuss the BV formalism, we have to relax this
ghost number restriction on $S(z)$. Namely, in order for the
antibracket and the delta-operator of eqs.~(\ref{eq:AB}) and
(\ref{eq:BOX}) to be non-vanishing and make sense, we have to allow
the existence in $S(z)$ of the couplings $s_n$ of any ghost number.
The situation is the same as in string field theory
\cite{HIKKO,Zwiebach-long}.

Once we relax the ghost number restriction (\ref{eq:restrictS}) on
$S(z)$, the same $I(S)$ as eq.~(\ref{eq:I}) satisfies the master
equation of TT:
\begin{equation}
\DELTA \invL I + \Half\AB{\invL I}{\invL I}=0 ,\label{eq:TTBV}
\end{equation}
for an arbitrary coupling constant $\lambda$:
namely, each term of eq.~(\ref{eq:TTBV}) vanishes separately:
\begin{eqnarray}
&&\AB{I}{I}= \int\! d\mu\,\Delta\B{H}\cdot\Delta H
= \int\! d\mu\,\B{H}\Delta^2 H = 0  ,\label{eq:ABII}\\
&&\DELTA I= \Half\mbox{Tr}\Delta
= \Half\int\prod_I dz^I \Delta\delta(z-z')\vert_{z'=z} = 0 .
\label{eq:DELTAI}
\end{eqnarray}
$\DELTA I$ vanishes because it is a Grassmann odd constant which we do
not have.\footnote{
In fact a more careful analysis using a suitable regularization may
prove necessary, and this might give a non-vanishing $S$ dependent
``anomaly'' to eq.~(\ref{eq:DELTAI}).}

\subsection{BV quantization of TT}

We now apply the BV quantization method to our TT described by the
actional $(1/\lambda)I(S)$. The expectation value of an observable
$\calO(S)$ is given by (cf. eq.~(\ref{eq:vevA}))
\begin{equation}
\VEV{\calO}=\frac{1}{\calZ(\lambda)}\int_\calL \calD\Lambda\,\calO(S)
\exp\left(\invL I(S)\right) ,\label{eq:VEV}
\end{equation}
where $\calL$ is a Lagrangian submanifold of the supermanifold of
$H(z)$, and $\calD\Lambda$ is the volume element on $\calL$ defined
similarly to eq.~(\ref{eq:dlambda}) on the basis of the volume element
$\calD H(z)$ on the total supermanifold of $H(z)$ (the explicit
expression for $\calD \Lambda$ will be given later).
$\calZ(\lambda)$ in eq.~(\ref{eq:VEV}) is the partition function of TT.
In order for the expectation value $\VEV{\calO}$ to be independent of the
choice of the Lagrangian submanifold $\calL$, the observable
$\calO(S)$ has to satisfy the condition (recall
eq.~(\ref{eq:observable}))
\begin{equation}
\DELTA \calO + \invL\AB{I}{\calO}=0 .\label{eq:Observable}
\end{equation}

As an example of an observable satisfying eq.~(\ref{eq:Observable})
we have the ``partition function operator'' $V_L(S)$:
\begin{equation}
V_L(S)=\int_L d\lambda\,H(z) .\label{eq:VL}
\end{equation}
For this $V_L$ each term of eq.~(\ref{eq:Observable}) vanishes
separately. $\DELTA V_L=0$ is obvious since $V_L$ is linear in $H(z)$.
As for the second term, $\AB{I}{V_L}$, we have
\begin{equation}
\AB{I}{V_L}= \int_L d\lambda\,\Delta H(z) = 0 ,\label{eq:ABIVL}
\end{equation}
where use has been made of the general property (\ref{eq:Schwarz-th})
of the integration over the Lagrangian submanifold $L$.

$\VEV{V_L}$ is not only independent of the choice of $\calL$ but in
fact is also independent of the choice of the Lagrangian submanifold $L$
defining $V_L$. This may be seen as follows. First, under a small
deformation $\delta_L$ of $L$ corresponding to the canonical
transformation $z^I\rightarrow z^I + \{z^I,\e(z)\}$, the observable
$V_L$ transforms as (cf. eq.~(\ref{eq:deltaL}))
\begin{equation}
\delta_LV_L(S)= \int_L d\lambda\Bigl(\Delta\e\cdot H + \{H,\e\}\Bigr)
=\int_L d\lambda\,\e\,\Delta H = \AB{X}{I} ,\label{eq:dLVL}
\end{equation}
with
\begin{equation}
X=\int_L d\lambda\,\e(z) H(z) .\label{eq:X}
\end{equation}
Then the variation of the expectation value $\VEV{V_L}$ under
$\delta_L$ is
\begin{eqnarray}
&\delta_L\VEV{V_L}&=\lambda\int_\calL\calD\Lambda\AB{X}{e^{I/\lambda}}
\nonumber\\
&&=\lambda\int_\calL\calD\Lambda\biggl(-\DELTA\!\left(Xe^{I/\lambda}\right)
+ \DELTA X\cdot e^{I/\lambda}-X\DELTA e^{I/\lambda}\biggr)\nonumber\\
&&=0 ,\label{eq:dLVEVVL}
\end{eqnarray}
where the three terms in the last expression vanish separately upon
using $i$) the formula (\ref{eq:Schwarz-th}) applied to TT,
$ii$) the fact that $X$ is linear in $H(z)$, and $iii$) the master
equation for $I(S)$, eq.~(\ref{eq:TTBV}) (in eq.~(\ref{eq:dLVEVVL}) we
have omitted to divide the RHSs by $\calZ(\lambda)$).

\subsection{Choosing a Lagrangian submanifold}

The quantum TT is now given by eq.~(\ref{eq:VEV}).
In this subsection we shall choose a concrete Lagrangian submanifold
$\calL$ for the quantization of TT.
For this purpose we shall work in the Darboux frame for $\calM$ with
$\rho(z)=1$ and treat one pair of field and antifield variables in a
special manner. We choose a frame
\begin{equation}
z^I=\left(\tau,\theta,\wt{z}^I\right),\qquad
\wt{z}^I=\left(\phi^i,\phi^*_i\right)_{i=1,\ldots,n-1} ,
\label{eq:wtz}
\end{equation}
with
\begin{equation}
\Delta=\DDiff{_l}{\tau}{\theta} + \wt{\Delta} \ ,\qquad
\wt{\Delta}=\sum_{i=1}^{n-1}(-)^{\phi^i}\DDiff{_l}{\phi^i}{\phi^*_i} ,
\label{eq:wtDelta}
\end{equation}
where $\tau$ and $\theta$ are Grassmann even and odd respectively,
and the condition $\Ngh[\tau]+\Ngh[\theta]=-1$ is satisfied.
Then we make explicit the dependence on the Grassmann odd coordinate
$\theta$ by expressing $H(z)$ as
\begin{equation}
H(z)=h(\tau,\wt{z}) + \theta \chi(\tau,\wt{z}) .\label{eq:h-chi}
\end{equation}
In terms of the components $h$ and $\chi$ of eq.~(\ref{eq:h-chi}), the
functional differentiation $\delta/\delta H(z)$ is given by
\begin{eqnarray}
&&\lFDiff{H(z)}=\theta\lFDiff{h(\tau,\wt{z})} +
\lFDiff{\chi(\tau,\wt{z})} ,\nonumber\\
&&\rFDiff{H(z)} = -\rFDiff{h(\tau,\wt{z})}\theta +
\rFDiff{\chi(\tau,\wt{z})} .\label{eq:FDiffH}
\end{eqnarray}
Note that $\delta/\delta h$ and $\delta/\delta\chi$ are Grassmann even
and odd, respectively.
Using eq.~(\ref{eq:FDiffH}), the antibracket (\ref{eq:AB}) of TT is
re-expressed as
\begin{eqnarray}
&&\AB{\calA}{\calB}=\int d\tau d\wt{z}\,
\calA\left(\rFDiff{h}\lFDiff{\B{\chi}}-\rFDiff{\chi}\lFDiff{\B{h}}
\right)\calB  ,\label{eq:ABhchi}
\end{eqnarray}
where $d\wt{z}\equiv\prod_id\phi^i d\phi^*_i$, and $h=h(\tau,\wt{z})$
and $\B{\chi}=\chi(\tau,\B{\wt{z}})$, etc.
Namely, we have chosen a Darboux coordinate for the supermanifold
of $H(z)$.

The Lagrangian submanifold $\calL$ in the space of functions $H(z)$
is specified by
\begin{equation}
\calL : \Gamma(\tau,\wt{z})=0 , \label{eq:Gammaeq0}
\end{equation}
where $\Gamma(\tau,\wt{z})$, which is defined for each
$(\tau,\wt{z})$, is a functional of $h$ and $\chi$, and it satisfies
the condition
\begin{equation}
\AB{\Gamma(\tau_1,\wt{z}_1)}{\Gamma(\tau_2,\wt{z}_2)}=0 ,
\label{eq:ABGammaGamma}
\end{equation}
for any pair $(\tau_1,\wt{z}_1)$ and $(\tau_2,\wt{z}_2)$.
A solution to eq.~(\ref{eq:ABGammaGamma}) is obtained by
assuming that
\begin{equation}
\Gamma(\tau,\wt{z}) = \chi(\tau,\wt{z}) + \widehat\Gamma(\tau,\wt{z}) ,
\end{equation}
where $\widehat\Gamma$ depends only on $h$.
Then eq.~(\ref{eq:ABGammaGamma}) is rewritten as
\begin{equation}
\AB{\Gamma(\tau_1,\wt{z}_1)}{\Gamma(\tau_2,\wt{z}_2)}=
\FDiff{\widehat\Gamma(\tau_1,\wt{z}_1)}{h(\tau_2,\B{\wt{z}}_2)}
- \FDiff{\widehat\Gamma(\tau_2,\wt{z}_2)}{h(\tau_1,\B{\wt{z}}_1)}=0 ,
\label{eq:ABGammaGamma2}
\end{equation}
and $\widehat\Gamma$ should be given as a gradient form
\begin{equation}
\widehat\Gamma(\tau,\wt{z})= \FDiff{G[h]}{h(\tau,\B{\wt{z}})}
\equiv\FDiff{G[h]}{\B{h}} ,\label{eq:Gammahat}
\end{equation}
in terms of the gauge fermion $G[h]$ of TT (cf. eq.~(\ref{eq:LinDB})).

\subsection{TT as a topological theory}

In this subsection we shall carry out a concrete study of the BV
quantized TT of eq.~(\ref{eq:VEV}) using the Lagrangian submanifold of
the previous subsection.
First, $I(S)$ and the pre-BRST transformation $\dB$ for TT
\begin{equation}
\dB H = \AB{H}{I} = \Delta H ,\label{eq:dB}
\end{equation}
are expressed in terms of $h$ and $\chi$ of eq.~(\ref{eq:h-chi}) as
\begin{equation}
I(S)=-\Half\int\! d\tau d\wt{z}\left(\B{\chi}\Diff{}{\tau}\chi +
\B{\chi}\wt{\Delta}h + \B{h}\wt{\Delta}\chi\right) , \label{eq:I-comp}
\end{equation}
and
\begin{eqnarray}
&&\dB h = \wt{\Delta}h + \Diff{\chi}{\tau} ,\nonumber\\
&&\dB\chi = \wt{\Delta}\chi .
\label{eq:dB-comp}
\end{eqnarray}
We shall consider the expansion around a classical solution $H_0(z)$
of the master equation:
\begin{equation}
H_0 = h_0 + \theta\chi_0 \quad \mbox{with}\quad \Delta H_0=0 .
\end{equation}
The master equation for the components $h_0$ and $\chi_0$ reads
\begin{eqnarray}
&&\wt{\Delta}h_0 + \Diff{\chi_0}{\tau} = 0 ,\nonumber\\
&&\wt{\Delta}\chi_0 = 0 .\label{eq:eqmot-comp}
\end{eqnarray}
Defining the fluctuation $f$ by
\begin{equation}
f(\tau,\wt{z})\equiv h(\tau,\wt{z})-h_0(\tau,\wt{z}) ,
\end{equation}
the Lagrangian submanifold $\calL$ is specified by
\begin{equation}
\calL : \chi(\tau,\wt{z})=\chi_0(\tau,\wt{z}) +
\FDiff{G[f]}{f(\tau,\B{\wt{z}})} .\label{eq:Lag}
\end{equation}
Note that this is slightly modified compared to the form in the
previous subsection because of the redefinition of $G$.
We assume that $\delta G[f]/\delta\B{f}\vert_{f=0}=0$ and therefore
$\chi\vert_{f=0}=\chi_0$ on $\calL$.

The gauge-fixed actional $\widehat I(f)$ and the corresponding BRST
transformation $\widehat \dB$ are given by
\begin{eqnarray}
&&\widehat I(f)\equiv I\vert_\calL = - \int\! d\tau d\wt{z}\left(
\Half\FDiff{G}{f}\cdot\Diff{}{\tau}\FDiff{G}{\B{f}}
+ \FDiff{G}{f}\cdot\wt{\Delta}f\right) ,\label{eq:Ihat}\\
&&\widehat\dB f \equiv \dB f\vert_\calL
=\wt{\Delta}f + \Diff{}{\tau}\FDiff{G}{\B{f}} ,\label{eq:dBhat}
\end{eqnarray}
where $\vert_\calL$ means the restriction to the Lagrangian
submanifold (\ref{eq:Lag}).
It can be checked explicitly that $\widehat I(f)$ is invariant under
$\widehat \dB$ and that the latter is on-shell nilpotent:
\begin{eqnarray}
&&\widehat \dB \widehat I(f) = 0 ,\label{eq:dBIeq0}\\
&&\left(\widehat \dB\right)^2\!f = \Diff{}{\tau}\left(
\FDiff{\widehat I}{\B{f}}\right) .\label{eq:dBsquare}
\end{eqnarray}
In proving eq.~(\ref{eq:dBIeq0}) we have used, in particular, the
following manipulation:
\begin{equation}
\int\! d\tau d\wt{z}\,\FDiff{G}{f}\,\wt{\Delta}\!
\left(\Diff{}{\tau}\FDiff{G}{\B{f}}\right)
=\int\! d\tau d\wt{z}\,\wt{\Delta}\!
\left(\Diff{}{\tau}\FDiff{G}{f}\right)\!\cdot\!\FDiff{G}{\B{f}}
= - \int d\tau d\wt{z}\wt{\Delta}\!
\left(\Diff{}{\tau}\FDiff{G}{\B{f}}\right)\!\cdot\!\FDiff{G}{f} = 0 ,
\end{equation}
where the first equality is due to partial integrations, and at the
second equality we have made a change of integration variables
$\wt{z}\rightarrow\B{\wt{z}}$, under which we have
$d\B{\wt{z}}=d\wt{z}$ and $\B{\wt{\Delta}}=-\wt{\Delta}$.

The partition function operator $V_L$ of eq.~(\ref{eq:VL}) on $\calL$
is given by
\begin{equation}
\widehat V_L(f) = V_L\vert_\calL=
V_L^0 + \int_L d\lambda\left(f+\theta\FDiff{G}{\B{f}}\right) ,
\label{eq:hatVL}
\end{equation}
where $V_L^0\equiv \int_L d\lambda H_0(z)$ is the partition function
of a gauge theory described by the action $S_0(z)=\ln H_0(z)$.
The expectation value $\VEV{V_L}$ of eq.~(\ref{eq:VEV}) may now be
explicitly written as
\begin{equation}
\VEV{V_L}=\frac{1}{\calZ(\lambda)}\int\!\calD f\,\widehat V_L(f)
\exp\left(\invL \widehat I(f)\right) .\label{eq:VEV2}
\end{equation}

Some comments are in order.
First, it should be noted that the range of path-integration
over $f$ in eq.~(\ref{eq:VEV2}) is non-trivial since the original
variable $H(z)$ is in fact an exponential function $H(z)=\exp S(z)$. It
is restricted to the region
\begin{equation}
f\vert_{\theta_i=0} > -h_0\vert_{\theta_i=0} ,\label{eq:range}
\end{equation}
where $\theta_i$ denotes all of the Grassmann odd coordinates in
$\wt{z}^I$. Therefore, even if we adopt $G[f]$ which is quadratic in
$f$ and hence makes $\widehat I$ (\ref{eq:Ihat}) quadratic in $f$, TT
is not truly a free field theory: $\widehat I$ expressed in terms of
unrestricted variable contains interactions.

Second, the (formal) invariance of the path-integral measure
$\calD f$ under the BRST transformation $\widehat \dB$ of
eq.~(\ref{eq:dBhat}) may be checked as follows:
\begin{equation}
\widehat \dB\ln\calD f = \int\! d\tau d\wt{z}
\FDiff{\left(\dB f(\tau,\wt{z})\right)}{f(\tau,\wt{z})}
= \mbox{Tr}\wt{\Delta} +
\int\! d\tau d\wt{z}\FDiff{}{f(\tau,\wt{z})}\cdot\Diff{}{\tau}
\FDiff{G}{f(\tau,\B{\wt{z}})} = 0 ,
\end{equation}
where the vanishing of the last term can be understood by partially
integrating with respect to $\tau$ and then changing the integration
variables from $\wt{z}$ to $\B{\wt{z}}$ to obtain minus the
original expression.

The third comment is that $\widehat V_L(f)$ (\ref{eq:hatVL})
is not exactly a $\widehat \dB$ invariant operator but
$\widehat\dB\widehat V_L$ is proportional to the equation of motion:
\begin{equation}
\widehat \dB \widehat V_L(f)=-\int_L d\lambda\,\theta\FDiff{\widehat
I(f)}{\B{f}} .\label{eq:hatdBhatVL}
\end{equation}
The invariance of $\VEV{V_L}$ under an infinitesimal change of the
gauge fermion $G$ by $\delta_G G$ can be checked explicitly using
eq.~(\ref{eq:hatdBhatVL}) and
$\delta_G \widehat I = \widehat\dB\left(\delta_G G\right)$, etc.

The gauge-fixed actional (\ref{eq:Ihat}) consists solely of terms
containing the gauge fermion $G[f]$ which specifies the gauge fixing.
Therefore one may suspect that our TT is a topological theory which
has no physical degrees of freedom as a quantum theory of $S$.
We show in the following that this is the case: by introducing an
auxiliary field the gauge-fixed actional $\widehat I$ can be reduced
to a BRST exact form.
For this purpose we multiply both the denominator and the numerator
of eq.~(\ref{eq:VEV2}) by a Gaussian integration over a new
Grassmann odd variable $B(\tau,\wt{z})$:
\begin{equation}
\int \calD B \exp\left\{
\frac{1}{2\lambda}\int\! d\tau d\wt{z}\left(\B{B}-\FDiff{G}{f}\right)
\!\Diff{}{\tau}\!\left(B-\FDiff{G}{\B{f}}\right)\right\} ,
\label{eq:GaussianInt}
\end{equation}
and consider the system of $f$ and $B$ variables (note that
(\ref{eq:GaussianInt}) is independent of the old variable $f$).
Let us define the new BRST transformation $\hhdB$ for the
$(f,B)$ system by
\begin{eqnarray}
&&\hhdB f = \wt{\Delta}f + \Diff{B}{\tau} ,\nonumber\\
&&\hhdB B = \wt{\Delta} B \label{eq:hhdB} .
\end{eqnarray}
This $\hhdB$ is apparently the same as the pre-BRST
transformation $\dB$ of eq.~(\ref{eq:dB-comp}) with $(h,\chi)$
replaced with $(f,B)$ and hence is off-shell nilpotent:
\begin{equation}
\left(\hhdB\right)^2 = 0 . \label{eq:off-shellNilpotent}
\end{equation}
The actional $\hhI(f,B)$ for the new $(f,B)$ system is obtained
by summing $\widehat I(f)$ (\ref{eq:Ihat}) and the contribution from
eq.~(\ref{eq:GaussianInt}), and it is written as a $\hhdB$ exact form:
\begin{eqnarray}
&\hhI(f,B)&= \widehat I(f)
+ \Half\int\! d\tau d\wt{z}\left(\B{B}-\FDiff{G}{f}\right)
\Diff{}{\tau}\left(B-\FDiff{G}{\B{f}}\right) \nonumber\\
&&=\int\! d\tau d\wt{z}\left(\Half\B{B}\Diff{}{\tau}B -
\FDiff{G}{f}\cdot\Diff{B}{\tau} - \FDiff{G}{f}\cdot\wt{\Delta}f\right)
\nonumber\\
&&= \hhdB\left(G[f]-\Half\int\! d\tau d\wt{z}\B{B}f\right) .
\label{eq:hhI}
\end{eqnarray}

In the $(f,B)$ system described by the actional $\hhI$, the partition
function operator $\widehat V_L$ (\ref{eq:hatVL}) is equivalent to a
new operator $\hhV_L$,
\begin{equation}
\hhV_L(f,B)= V_L^0 + \int_L d\lambda\left(f + \theta B\right) ,
\label{eq:hhVL}
\end{equation}
so long as we consider only the the ``one-point'' function:
\begin{equation}
\VEV{V_L}=\frac{1}{\calZ(\lambda)}\int\!\calD f\!\int\!
\calD B\,\hhV_L(f,B)\exp\left(\invL\hhI(f,B)\right) .
\label{eq:VEVhhVL}
\end{equation}
This is because we have
\begin{equation}
\widehat V_L(f)=\hhV_L(f,B) - \int_L\! d\lambda\,\theta
\left(B-\FDiff{G}{\B{f}}\right) ,\label{eq:hatVLhhVL}
\end{equation}
and the expectation value of the last term of eq.~(\ref{eq:hatVLhhVL})
vanishes since it is odd with respect to $B-\delta G/\delta
\B{f}$ (cf. eq.~(\ref{eq:GaussianInt})).\footnote{
Here we are assuming that the discrete symmetry
$B\rightarrow -B+2\delta G/\delta\B{f}$ is not spontaneously broken.}
For a general multi-$V_L$ function $\VEV{\prod_i V_{L_i}}$ (whose
physical meaning is not clear at present), $\widehat V_L$ cannot be
replaced by $\hhV_L$.
In distinction to the case of the $\widehat V_L$ operator in the
$f$-formalism (cf., eq.~(\ref{eq:hatdBhatVL})), the $\hhV_L$ operator
in the $(f,B)$ formulation is simply a $\hhdB$ invariant operator:
\begin{equation}
\hhdB \hhV_L(f,B)=\int_L d\lambda\,\Delta\!\left(f+\theta B\right)=0 .
\label{eq:hhdBhhVL}
\end{equation}
This implies that $\VEV{V_L}$ is in fact independent of the coupling
constant $\lambda$:
\begin{equation}
\Diff{}{\lambda}\VEV{V_L}=0 ,\label{eq:VLislambdaindep}
\end{equation}
and therefore $\VEV{V_L}$ may be calculated in the ``classical'' limit
$\lambda\rightarrow 0$ \cite{Witten-TFT,BBRT}.
In this limit the $(f,B)$ fields are frozen,
and we find that the expectation value of the operator $V_L$ in TT is
equal to the partition function of the gauge theory corresponding to a
classical solution $S_0(z)$ of the master equation (\ref{eq:BV}):
\begin{equation}
\VEV{V_L}=\lim_{\lambda\rightarrow 0}\VEV{V_L}=V_L^0 .
\label{eq:calcVEVVL}
\end{equation}
This implies the equivalence between our TT and the ordinary
formulation of gauge theories, at least as far as the partition
function is concerned.

In deriving eq.~(\ref{eq:calcVEVVL}), we have implicitly assumed that
there are no stationary points of $\hhI(f,B)$ other than the trivial
one $(f,B)=(0,0)$.
What will happen if there are other stationary points of $\hhI$?
In this case there would be two possibilities: one is that we have to
sum all the contributions from the stationary points of $\hhI$, and
the other is that only one of the stationary points is chosen in the
same manner as in the case of spontaneous symmetry breakdown.
Which of the two is realized depends on the dynamics of TT: whether TT
with a non-vanishing coupling constant $\lambda$ is in a disordered
phase or in a ordered perturbative phase.

If all the stationary points of $\hhI$ have to be summed over, TT
would be a very peculiar theory which could hardly reproduce an
ordinary gauge theory.
To discuss the latter case of choosing one stationary point, let us
regard the variable $f$ as the original $h$ of eq.~(\ref{eq:h-chi}),
that is, let us put $H_0=0$ in the above equations. For simplicity we
consider the stationary points of $\widehat I(h)$ instead of those of
$\hhI(h,B)$ (they give the same result for $h$).
The stationary condition of $\widehat I(h)$ reads
\begin{equation}
\FDiff{\widehat I(h)}{h}= \left(\wt{\Delta} +
\widehat\dB\right)\FDiff{G[h]}{h}=0 . \label{eq:FDiffhIh}
\end{equation}
{}From eqs.~(\ref{eq:FDiffhIh}), (\ref{eq:eqmot-comp}) and
(\ref{eq:dBhat}) we see that, if we have a solution $H_0$ of the
master equation $\Delta H_0=0$ on the Lagrangian submanifold $\calL$
of the form
\begin{equation}
H_0 = h_0 + \theta \FDiff{G[h]}{h}\bigg\vert_{h=h_0} ,
\label{eq:solBVonL}
\end{equation}
then $h=h_0$ is automatically a stationary point of $\widehat I(h)$.
However, not all solutions of eq.~(\ref{eq:FDiffhIh}) correspond to
solutions of the master equation. If only one stationary point is
chosen in evaluating $\VEV{V_L}$ in TT, a stationary point $h_0$ which
does not correspond to a solution of the master equation should not be
selected, since in that case we would have
$\widehat\dB h\vert_{h=h_0}\neq 0$ and the BRST symmetry in TT would
be spontaneously broken so that the above argument leading to the
$\lambda$-independence of $\VEV{V_L}$ would break down. If we have
many stationary points which have corresponding solutions
(\ref{eq:solBVonL}) of the master equation, we do not yet know how one
of them might be singled out. Note that $\widehat I(h)$ vanishes at
every of these BRST invariant stationary points.

\section{Summary and discussion}
\reseteqnum

In this paper we have proposed a new approach to gauge theory.
It is formulated as a gauge theory having the action $S$ of ordinary
gauge theories as its dynamical variable.
Applying the BV quantization method, we have found that our TT is
essentially a topological theory described by a BRST exact actional
and can reproduce an ordinary gauge theory corresponding to an action
satisfying the master equation (\ref{eq:mastereq}).
Since this new formulation does not refer to any concrete solution of
the master equation (\ref{eq:mastereq}) in its basic formulation, our
TT is expected to be useful in its application to closed string field
theory whose quantum action takes a very complicated form.

There are many questions left unanswered. Most of them are mentioned
in the text. We finish this paper by summarizing them (not necessarily
in the order of importance).

\noindent {\em i}) Observables in TT. At present we have only the
partition function operator (\ref{eq:VL}) as an example of an
observable satisfying the condition (\ref{eq:Observable}).
In order to make clearer the connection between the present TT and the
ordinary formulation of gauge theory, we have to prepare more
observables. An ``on-shell amplitude operator'' would be an
interesting candidate, however, we do not know the explicit form of
such an operator in TT.

In relation to the problem of observables in TT, we comment that the
product of the partition function operators, $\calO=\prod_i V_{L_i}$,
is also an observable satisfying the condition (\ref{eq:Observable}).
We have $\AB{I}{\calO}=\sum_i\AB{I}{V_{L_i}}\prod_{j\neq i}V_{L_j}=0$
since $\AB{I}{V_{L_i}}=0$. As for $\DELTA \calO$, it is expressed using
eq.~(\ref{eq:Witten-formula}) as a sum of terms which contain either
$\DELTA V_{L_i}(=0)$ or
\begin{equation}
\AB{V_{L_i}}{V_{L_j}}=\int_{L_i}\! d\lambda(z)\int_{L_j}\! d\lambda(w)
\frac{1}{\rho(z)}\delta(z-\B{w}) .
\label{eq:ABVLVL}
\end{equation}
The quantity (\ref{eq:ABVLVL}) vanishes since it is a constant with
$\Ngh=1$. However, we do not know whether there is any interesting
meaning to the expectation value $\VEV{\prod_i V_{L_i}}$. Note that
$(\p/\p\lambda)\VEV{\prod_{i=1}^{N}V_{L_i}}=0$ does not hold
for $N\ge 2$.

\noindent {\em ii}) $\Ngh[I(S)]=0$?
We have left unanswered the question of whether the actional $I(S)$ of
eq.~(\ref{eq:I}) carries no ghost number, that is, whether the
condition (\ref{eq:Nghdmu}) is satisfied in the system we are
interested in, for example closed string field theory. As stated in
the text, this is not an easy problem since the index $I$ is in fact a
continuous parameter.

\noindent {\em iii})
We do not yet know how to treat the case where the actional $\hhI$ has
many stationary points, should this arise.

\noindent {\em iv})
It is an interesting question whether the present TT formulation
applied to closed string field theory gives a space-time background
independent formulation (see refs.~\cite{Witten-BI,SenZwiebach} for
recent studies on the background independence of string field theory).

\noindent {\em v})
We have no intuitive understanding of why TT, which has a nontrivial
classical actional, $I(S)$ of eq.~(\ref{eq:I}) with the restriction
(\ref{eq:restrictS}), is reduced upon BV quantization to a
topological theory, at least in the evaluation of $\VEV{V_L}$.

\noindent {\em vi})
Most of the manipulations in this paper are very formal. It is
desirable to study the validity of the arguments for TT using a simple
model.

The last and the most important problem is how the present TT
formulation is useful in the study of gauge theories, and in particular
of string theory. This problem is currently under investigation.

\vskip1cm
\centerline{\large\bf Acknowledgements}

I would like to thank K.-I.~Izawa for valuable discussions.
I also wish to acknowledge M.G.~Mitchard for careful reading of the
manuscript.

\newcommand{\J}[4]{{\sl #1} {\bf #2} (19#3) #4}
\newcommand{\MPL}{Mod.~Phys.~Lett.}
\newcommand{\NP}{Nucl.~Phys.}
\newcommand{\PL}{Phys.~Lett.}
\newcommand{\PR}{Phys.~Rev.}
\newcommand{\PRL}{Phys.~Rev.~Lett.}
\newcommand{\AP}{Ann.~Phys.}
\newcommand{\CMP}{Commun.~Math.~Phys.}


\end{document}